\documentclass[10pt,twocolumn,twoside]{IEEEtran}

\ifCLASSINFOpdf
  
\else
  
\fi

\newtheorem{rem}{Remark}
\newtheorem{defn}{Definition}
\newtheorem{lem}{Lemma}
\newtheorem{prop}{Proposition}

\newtheorem{thm}{Theorem}
\newtheorem{assm}{Assumption}

\newtheorem{problem}{Problem}

\usepackage{algorithm,algpseudocode}
\usepackage{cite}
\usepackage{amsmath}
\usepackage{caption}
\usepackage{amstext}
\usepackage{amssymb}
\usepackage{tikz}
\usepackage[mathscr]{euscript}
\usepackage{color}
\usepackage{amsfonts}
\usepackage{amsmath}
\usepackage{arydshln}

\usepackage[final]{pdfpages}
\usepackage{csquotes}
\makeatletter
\newcommand*{\rom}[1]{\expandafter\@slowromancap\romannumeral #1@}
\makeatother
\usepackage{graphicx}      

\usepackage{xparse}
\usepackage{mathdots}
\usepackage{epstopdf}

\newcommand{\seb}[1]{%
{\leavevmode\color{black}#1}%
}

\newcommand{\jezd}[1]{%
{\leavevmode\color{black}#1}%
}

\DeclareMathAlphabet\mathbfcal{OMS}{cmsy}{b}{n}
\usepackage{array}

\usepackage{comment}

\usepackage{mathtools,xparse}

\usepackage{pdfpages}
\usepackage{xcolor}
\usepackage{subcaption}
\usepackage{enumitem}

\DeclareMathOperator{\rank}{rank}

\DeclareMathOperator{\nrank}{normal-rank}
\DeclareMathOperator{\gnrank}{g-n-rank}
\begin{document}
\author{Sebin~Gracy,
        ~Jezdimir~Milo\v sevi\' c
	   ~and~Henrik~Sandberg
\thanks{The authors are with the Division of Decision and Control Systems, School of Electrical Engineering and Computer Science, KTH Royal Institute of Technology, Stockholm, Sweden. Emails: {gracy,jezdimir,hsan}@kth.se. }}
\title{Actuator Security Index for Structured Systems  }
\maketitle
\begin{abstract}
 Given a network with the set of vulnerable actuators (and sensors), the security index of an actuator equals the minimum number of sensors and actuators that needs to be compromised  so as to conduct a perfectly undetectable attack using the said actuator. This paper deals with the problem of computing actuator security indices for discrete-time LTI network systems. Firstly, we show that, under a structured systems framework, the actuator security index is generic. Thereafter, we provide graph-theoretic conditions for computing the structural actuator security index. The said conditions are in terms of existence of linkings on appropriately-defined directed (sub)graphs. Based on these conditions, we present an algorithm for computing the structural index.
\end{abstract}

\section{Introduction}


In recent years, security of cyber-physical systems   has attracted a lot of interest within the control systems community~\cite{6016202,7935369}. A large number of problems, previously considered in an attack-free setting, have been extended to account for cyber-attacks; examples include design of attack resilient estimators~\cite{tabuada_estimation}, detectors~\cite{pasqualetti_detection}, and consensus protocols~\cite{sundaram_consensus}.  However, new problems such as cyber-attacks modeling~\cite{andre_modeling}, attack analysis~\cite{Guo_linearattacks}, and development of security indices~\cite{jezdimir2019} have also emerged.  The present paper is concerned with the latter.

The security index, $\delta(i)$, is defined for every actuator $i$, and is equal to the minimum number of sensors and actuators that needs to be compromised by an attacker so as to conduct a perfectly undetectable attack using $i$. A large (resp. small) security index means that a large (resp. small) number of components have to be attacked in conjunction with the actuator of interest. Thus, the security index helps the system operator to characterize which actuators are the most vulnerable in the system, and can also be used to prioritize allocation of security investment~\cite{secalo3}.  

The notion of security index in dynamical network systems was first introduced  in \cite{sandberg2016control}. In the said paper, the security index was defined based on the notion of \emph{undetectable attacks}~\cite{pasqualetti_detection} i.e., attacks where the output corresponds to that generated by a disturbance, thus leading the system operator to conclude (erroneously) that the system is affected by disturbance as opposed to an attack. Yet another class of attacks is  that of \emph{perfectly undetectable attacks}~\cite{weerakkody2016graph}, i.e., attacks that do not leave a trace at the output. Clearly, perfectly undetectable attacks could lead to more serious consequences. Hence, from a security viewpoint, it is more prudent to define the security index with respect to perfectly undetectable attacks. As such, a novel (actuator) security index has been recently proposed in \cite{jezdimir2019}. Note that the said index is defined with respect to a given plant model, i.e., each of the system matrices having specific numerical entries. \jezd{Since variations of system parameters can result in actuator security indices changing their values, a robust security index, $\delta_{r}(i)$, has been proposed and developed based on  structural model of the system \cite{jezdimir2019}. However, this index is developed for finding only the most vulnerable actuators in the system, that is, those actuators that are vulnerable for \emph{every} admissible realization of the system parameters. Yet, some actuators can be vulnerable for \emph{almost all}, but not necessarily for \emph{every} realization of the system parameters. Hence, we aim to develop an index that is less strict than $\delta_{r}(i)$, yet still robust to numerical changes in modeling parameters. }



We consider a structured systems framework. That is, the system matrices are \emph{structured} matrices; certain positions are fixed to zero, while the positions that are not  a priori fixed to zero are referred to as \emph{free parameters}. Each numerical choice of free parameters yields a \emph{realization} of the system at hand. Under such a setting, our contributions are twofold: \jezd{First, we show that for a given actuator $i$, the security index $\delta(i)$ is \emph{generic} (Theorem~\ref{ref:main-result1}). That is, the security index of an actuator $i$ has the same value for almost all choices of free parameters, say $p$. We adopt this value to be the generic security index of an actuator $i$, and refer to it as $\delta_{s}(i)$.  Second, we provide graph-theoretic conditions for computing the generic security index (Theorem~\ref{ref:main-result2}), and,  based on those, an algorithm for the same (Algorithm~\ref{algorithm:impact-calculation}).}



We conclude this section by presenting the list of notations that would be used in the sequel. Thereafter, we formally state our problem of interest in Section~\ref{sect:propb:form}, while the necessary background material is provided in Section~\ref{sect:prelim}. The main results are provided in Section~\ref{sect:main:result}. Finally, we conclude with some closing remarks in Section~\ref{sect:conclusion}.  
\subsection*{Notations} $\mathbb{R}$,  $\mathbb{C}$ $\mathbb{Z}$, $\mathbb{Z}_{+}$ and $\mathbb{Z}_{\geq 0}$ denote the set of real numbers, complex numbers, integers, positive integers, and non-negative integers, respectively.    $I_{N}$ indicates an identity matrix of size $N$.    $\left|\mathcal{X}\right|$ indicates cardinality of a set $\mathcal{X}$.  For a  column vector $a$, $a^{(i)}$ denotes the element corresponding to its $i^{th}$ row.  

\section{Problem Formulation} \label{sect:propb:form}
\jezd{\subsection{Plant Model}
We  consider a linear time-invariant plant model
\begin{equation}
\begin{aligned}
x(k+1)&=Wx(k)+B u(k),\\
y(k)&=C x(k)
\label{eqn:system:plant:model}
\end{aligned}
\end{equation}
where:  
$x(k) \in \mathbb{R}^{N}$ is the state vector; 
$u(k) \in \mathbb{R}^{Q}$ is the control input; 
and $y(k) \in \mathbb{R}^{M}$ are the sensor measurements.}

\subsection{Attacker Model}
 Following~\cite{jezdimir2019},
we model the plant dynamics under attack as
\begin{equation}
\begin{aligned}
x(k+1)&=Wx(k)+B u(k) + B_{a}a(k),\\
y(k)&=C x(k) + D_{a}a(k)
\label{eqn:system:attacker:model}
\end{aligned}
\end{equation}
where $a(k)$$ \in$$ \mathbb{R}^{P}$ is the attack vector. 
Some sensors may be immune to attacks, in which case they are referred to as \emph{secure} sensors. 
We define the joint set of actuators and unprotected sensors by $\mathcal{I} =\{1,2, \hdots P\}$, the set of protected sensors by $f$, and  attack cardinality by
\begin{equation}
|a|_{0}=|\cup_{k \in \mathbb{Z}_{\geq 0}} supp(a(k))|, \nonumber \end{equation} where $\text{supp}(a(k))=\{i\in \mathcal{I}:a^{(i)}(k) \neq 0\}$.  \\
Note that the system~\eqref{eqn:system:plant:model} could be subject to actuator attacks as well as sensor attacks.
The first $Q<P$ elements of $a(k)$ represent  attacks on the actuator(s), while the remaining elements model (dedicated) attacks on the sensor(s). 
Consequently, $B_{a}$ and $D_{a}$ are as follows:
\[
  B_{a}=
  \begin{bmatrix} B && 0_{N \times (P-Q)}\end{bmatrix},\quad
  D_{a}=
 \begin{bmatrix} 0 & I_{M-|f|} \\ 0 & 0 \end{bmatrix}
\]

Note that it is a standard practice in the literature to assume dedicated attacks on the actuators, see for instance \cite{jezdimir2019,weerakkody2016graph}. Given that we make no assumptions on the input matrix, our results, in particular, apply to the case where the actuator attacks are dedicated.

  We are concerned with  \emph{perfectly undetectable attacks}~\cite{weerakkody2016graph} i.e., the attacker would like to conduct an attack without being detected by the system operator. These attacks are formally defined as follows.
\begin{defn} \label{defn:perfect}
Let $y(x(0), u,a)$ denote the response of system~\eqref{eqn:system:attacker:model} for the initial condition $x(0)$, input $u$, and attack signal $a \neq 0$. Then, the attack signal $a$ is perfectly undetectable if $y(x(0),u,a) = y(x(0), u,0)$.~$\blacksquare$
\end{defn}
Since system~\eqref{eqn:system:attacker:model} is linear, we can set $x(0) =0$ and $u =0$ without any loss of generality.

 Based on this notion of undetectability,  ~\cite{jezdimir2019} introduces the actuator security index.
Recall that this index quantifies the minimum number of sensors and actuators that needs to be compromised by the attacker to attack actuator $i$ and remain perfectly undetectable simultaneously.  Hence, for an actuator $i$, the problem of computing $\delta(i)$  can then be defined in the following way.

\begin{problem}\label{problem:sec_index_perfectly_undetectable} \textit{Calculating $\delta$}
\begin{align*}
\delta(i):= &\underset{a}{\text{minimize}} &&||a||_{0} &&&\hspace{1mm}\\
&\text{subject to}   &&x(k+1)=Wx(k)+B_a a(k), &&& \text{(C1)}\\
    &\hspace{5mm}    &&0=Cx(k)+D_a a(k),   &&&\text{(C2)} \\
   &\hspace{5mm}     &&x(0)=0, 
   &&&\text{(C3)}\\
    &\hspace{5mm}    &&a^{(i)} \neq 0.  &&&\text{(C4)}
\end{align*}
\end{problem}
\seb{
In words, the optimal solution of the problem equals the minimum number of sensors and actuators to conduct a perfectly undetectable attack. 
The constraints (C1) and (C2) ensure that the physical dynamics is according to the model we introduced, while  (C2) and (C3) ensure that the attack signal $a$ is perfectly undetectable. 
Finally, since we are calculating the security index for actuator $i$, by definition we need to ensure that actuator $i$ is actively used in the attack. 
This is accomplished by introducing constraint (C4).} 
As pointed out in \cite{jezdimir2019} it is not always the case that Problem~\ref{problem:sec_index_perfectly_undetectable} has a solution. If, for a given actuator $i \in \mathcal{I}$, Problem~\ref{problem:sec_index_perfectly_undetectable} does not have a solution, then we adopt the notation: $\delta(i) = + \infty$.

\color{black}

Given that we are interested in understanding how $\delta(i)$ changes with respect to variations in system matrices, we turn our attention to structured systems.

\subsection{Structured System and Graphical Representation}
Inspired by the graphical representations of structured systems in  the literature (see for instance \cite{FCDC17,weerakkody2016graph}), we introduce graphical representations of \eqref{eqn:system:plant:model} and \eqref{eqn:system:attacker:model} in this subsection. \\

Let $W_{\omega}$, $B_{\beta}$ and $C_{\gamma}$ be structured matrices. That is, they have positions that are a priori fixed to zero, and the ones that are not fixed to zero are referred to as free parameters. The interpretation is in the following sense: the fixed zero positions represent interactions that are prohibited from occurring, while the free parameters denote interactions that may happen without specifying how intense  these interactions may be. 
Let $s_{1}$, $s_{2}$ and $s_{3}$ denote the number of free parameters in $W_{\omega}$, $B_{\beta}$ and $C_{\gamma}$, respectively, and let $\mathbb{R}^{\alpha_{1}}$, where $\alpha_{1} =  s_{1} + s_{2} + s_{3}$, denote the space of free parameters. We define  $\omega \in \mathbb{R}^{s_{1}}$ to be the vector of free parameters in $W_{\omega}$.  Analogously, we define $\beta$ and $\gamma$. Each choice of free parameter in $\mathbb{R}^{\alpha_{1}}$ yields a system $(W, B, C)$ having dynamics as given in \eqref{eqn:system:plant:model}.

Let $X =\{x_{1}, x_{2}, \hdots, x_{N}\}$,  $Y =\{y_{1}, y_{2}, \hdots, y_{M}\}$, and $U =\{u_{1}, u_{2}, \hdots, u_{Q}\}$ denote the set of state vertices, output vertices and actuator vertices, respectively. We define the associated edge sets as follows: $\mathcal{E}_{W} =\{(x_{j}, x_{i}) \subseteq X \times X\mid [W_{\omega}]_{ij} \neq 0)\}$,   $\mathcal{E}_{B} =\{(u_{j}, x_{i})  \subseteq U \times X\mid [B_\beta]_{ij} \neq 0)\}$, $\mathcal{E}_{C} =\{(x_{j}, y_{i}) \subseteq X \times Y\mid [C_\gamma]_{ij} \neq 0)\}$. Let $\mathcal{G}=(\mathcal{V}, \mathcal{E})$, where $\mathcal{V} = X  \cup U \cup Y$ and $\mathcal{E} = \mathcal{E}_{W}  \cup \mathcal{E}_{B} \cup \mathcal{E}_{C}$, be the graph associated with the system \eqref{eqn:system:plant:model}.

 We define $B_{\beta^{\prime}}$ and $D_{\sigma}$ as  structured matrices whose columns correspond to vertices of $\mathcal{I}$ and rows correspond to vertices of $X$ and $Y$, respectively. Let $B_{\beta^{\prime}} = [\begin{smallmatrix} B_{\beta} & 0\end{smallmatrix}]$. Therefore, the number of free parameters in $B_{\beta^{\prime}}$ is the same as that in $B_{\beta}$, and, hence, equal to $s_{2}$. Let $s_{4}$ be the number of free parameters in $D_{\sigma}$, and notice that the edges in $\mathcal{E}_{Y, Y_{F}}$ are in one to one correspondence with the free parameters in $D_{\sigma}$.  Let $\mathbb{R}^{\alpha}$, where $\alpha = \alpha_{1} + s_{4}$, denote the space of free parameters for the structured system under attack. Each choice of free parameters in $\mathbb{R}^{\alpha}$ yields a system $(W, B_{a}, C, D_{a})$ having dynamics as given in \eqref{eqn:system:attacker:model}.

\begin{figure}[h!]
\centering
\includegraphics[height=0.30\columnwidth]{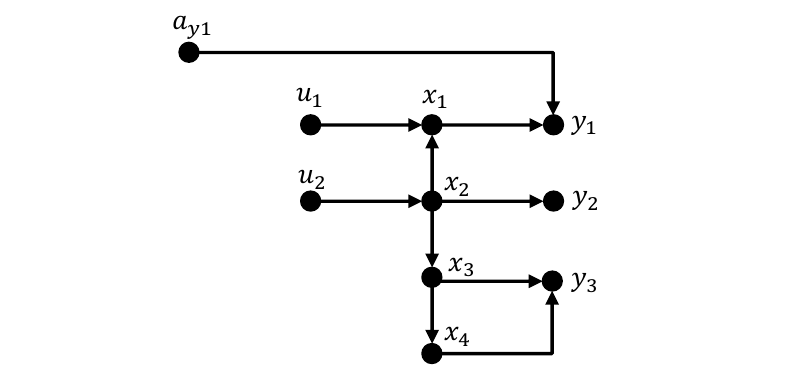}
\caption{Graph $\mathcal{G}_{1}$}

\label{fig:running:example}
\end{figure}

Recall that $\mathcal{I}$ denotes the joint set of actuators and sensors that are vulnerable to attacks. With respect to $\mathcal{G}$, $\mathcal{I} =\{u_{1}, u_{2}, \hdots, u_{Q}, y_{Q+1}, \hdots, y_{Q+P-Q}\}$, where $u_{i} \in U$ and $y_{i} \in Y$. While no assumptions are made on actuator attacks, we do assume that the sensor attacks are dedicated. Hence,  we introduce  corresponding set of (dedicated) sensor attack nodes $Y_{F} = \{a_{y_{\ell}}, a_{y_{\ell +1}}, \hdots, a_{y_{P-Q}}\}$, and the associated edge set    $\mathcal{E}_{D_{a}} =\{(y_{i}, a_{y_{j}}) \subseteq Y_{F} \times Y \mid [D_{\sigma}]_{ij} \neq 0\}$. Since the attacks are dedicated,  $[D_{\sigma}]_{ij} \neq 0$ if and only if i) $j > Q$  and ii) $i=j$.   Let  $\mathcal{G}_{F} = (\mathcal{V}^\prime, \mathcal{E}^\prime)$, where $\mathcal{V}^\prime = V  \cup Y_{F}$ and $\mathcal{E}^\prime = \mathcal{E} \cup  \mathcal{E}_{D_{a}}$, be the graph associated with the system \eqref{eqn:system:attacker:model}. An illustration of this setup is given in Figure~\ref{fig:running:example}, which will be the running example in this paper. 

 Note that the representation in terms of $\mathcal{G}_{F}$ (resp. $\mathcal{G}$) and structured matrices $(W_{\omega}, B_{\beta^{\prime}}, C_{\gamma}, D_{\sigma})$ (resp. $(W_{\omega}, B_{\beta}, C_{\gamma})$) are in one-to-one correspondence. Hence throughout this paper, we use the terms \enquote{choice of edge weights} and \enquote{realization} interchangeably as the context warrants.

 So as to avoid degenerate cases, we make the following assumption.
\begin{assm}\label{assm:dummy}
\begin{enumerate}[label=(\roman*)]
\item Every actuator $u_{i} \in U$ acts upon at least one state $x_{j} \in X$.
\item Every state vertex $x_{i} \in X$ has a path to at least one vertex $y_{k} \in Y$.
\end{enumerate}
\end{assm}

%
 
\subsection{Problem Statement}\label{pb:stmnt}

\jezd{The main objective of the present paper is to answer the following: 
\begin{enumerate}[label=(\roman*)]
\item \label{q1} For a given actuator, $i \in \mathcal{I}$, does $\delta(i)$  have the same value for almost all
choices of edge weights in $\mathcal{G}_{F}$? 
\item \label{q2} Secondly, assming that the aforementioned question has a positive answer, how does one \emph{compute} that value? 
\end{enumerate}
 \noindent Hereafter, for a given $i \in \mathcal{I}$, we use \enquote{$\delta_{s}(i)$} in place of \enquote{$\delta(i)$ for almost all choices of edge weights in $\mathcal{G}_{F}$}.
 }

\section{Preliminaries} \label{sect:prelim}


\subsection{Structured matrices and associated ranks}
We recall some notions related to ranks in this subsection.
The matrix pencil associated with a realization $(W, B_{a}, C, D_{a})$ is defined by \begin{equation}P(z) = \begin{bmatrix}W -zI_{N} & B_{a}\\C& D_{a}  \end{bmatrix}.\nonumber \end{equation}
The normal rank of $P(z)$ is the maximum rank that $P(z)$ attains for some $z_{0} \in \mathbb{C}$.


Define \begin{equation}P_{\alpha}(z) = \begin{bmatrix}W_{\omega}-zI_{N} & B_{\beta^{\prime}}\\C_{\gamma} & D_{\sigma}  \end{bmatrix}.\nonumber \end{equation}Note that for each numerical choice of free parameters in $\mathbb{R}^{\alpha}$, we obtain the matrix pencil, $P(z)$, associated with the corresponding realization $(W, B_{a}, C, D_{a})$. Subsequently, we can compute the normal rank for such a $P(z)$. It turns out that $\nrank(P(z))$ is the same value for almost all choices of free parameters in $\mathbb{R}^{\alpha}$ \cite{van1991structure}. Below we recall a formal definition.
\begin{defn}\label{defn:genericnrank}[Defintion C.3 \cite{sundaram2012fault}] The generic normal rank of $P_{\alpha}(z)$ ($\gnrank P_{\alpha}(z)$) is the maximum rank that a matrix pencil $P(z)$ achieves over all choices of free parameters in $\mathbb{R}^{\alpha}$ and $z \in \mathbb{C}$.~$\square$
\end{defn}

\subsection{Graph Vocabulary}

The advantage of a structured systems approach is that, thanks to the fixed zero pattern of the structured system matrices, such systems can be represented using graphs. Consequently, they can be studied using tools from graph-theory. In this subsection, we familiarize ourselves with some graph-theoretic notions, which would be used in the sequel (see \cite{godsil2001algebraic} for  more detailed treatment). 


Consider a graph $\mathcal{G}=(\mathcal{V}, \mathcal{E})$. Let $w_{p}$, $w_{p+r}$ $\in \mathcal{V}$. A \emph{path} from $w_{p}$ to $w_{p+r}$ in $\mathcal{G}$  is a sequence of edges $w_{p} \rightarrow w_{p+1}$, $w_{p+1} \rightarrow w_{p+2}$, $\hdots$, $w_{p+r-1} \rightarrow w_{p+r}$. Two paths are said to be \emph{vertex-disjoint} if they have no vertex in common. 
\begin{defn}\label{defn:linking} Let $S_1$ and $S_2$ be two sets of vertices of a directed graph. A collection of vertex-disjoint paths from $S_1$ to $S_2$ is called a \emph{linking} from $S_1$ to $S_2$.~$\blacksquare$
\end{defn} 
Suppose $L$ is a linking in $\mathcal{G}$. If $q$ is one of the vertices along the paths in $L$, we say that $L$ \emph{saturates} $q$. The size of a linking is the \emph{number} of paths contained in the linking. Let $S_1$ and $S_2$ be two sets of vertices in $\mathcal{G}$. If a linking, $L$, from $S_1$ to $S_2$ has maximum size among all linkings from $S_1$ to $S_2$, then $L$ is a \emph{maximum $S_1$-$S_2$ linking} in $\mathcal{G}$.   Note that maximum linkings are not necessarily unique. Vertices that are present in every maximum linking from $S_1$ to $S_{2}$ are referred to as \emph{essential} vertices between vertex sets  $S_{1}$ and $S_{2}$, and denoted as $V_{ess}(S_{1}, S_{2})$.

Indeed, the maximum size of a linking on a directed graph relates to the generic normal rank of the corresponding structured transfer function matrix. We recall this connection in the following lemma.

\begin{lem}[\cite{van1999generic}, Pg.140 \cite{sundaram2012fault}] \label{lem:vanderwoude}
The maximum number of vertex-disjoint paths from the set of input vertices to the set of output vertices in $\mathcal{G}_{F}$ equals $\gnrank C_{\gamma}(zI-W_{\omega})^{-1}B_{\beta^{\prime}} + D_{\sigma}$.~$\blacksquare$
\end{lem}

\section{Main Result} \label{sect:main:result}
 In this section, we establish the main contributions of the present paper. Our first main result is Theorem~\ref{ref:main-result1}, where we show that the actuator security index is generic.  Our second main result is Theorem~\ref{ref:main-result2}, wherein we provide graph-theoretic conditions for computing the generic actuator security index.


 Let $\mathcal{I}_{a} \subseteq \mathcal{I}$ denote the (sub)set of actuators and unprotected sensors that are used in an attack.  Let us denote by $(W, \tilde{B}_{a}, C, \tilde{D}_{a})$ the corresponding system, where $[\begin{smallmatrix} \tilde{B}_{a} \\ \tilde{D}_{a}\end{smallmatrix}]$ the submatrix of $[\begin{smallmatrix} B_{a} \\ D_{a}\end{smallmatrix}]$ formed by looking at columns of  $[\begin{smallmatrix} B_{a} \\ D_{a}\end{smallmatrix}]$ that correspond to elements in $\mathcal{I}_{a}$. Denote by $G_{\alpha}^{\mathcal{I}_{a}}(z)$ the corresponding structured  transfer function matrix.\\Before we proceed to our first result, we need the following intermediate results:


\begin{lem} \label{lem:every:max:linking} Let $\mathcal{I}_{a} \subseteq \mathcal{I}$. If for some $i \in \mathcal{I}_{a}$, $\gnrank G^{\mathcal{I}_{a}} = \gnrank G^{\mathcal{I}_{a}\setminus i} +1$, then vertex $i$ is saturated by every maximum linking from $\mathcal{I}_{a}$ to $Y$.~$\blacksquare$
\end{lem}
\textit{Proof:} Let $\gnrank  G^{\mathcal{I}_{a}} =r$, where $1 \leq r \leq \lvert\mathcal{I}_{a}\rvert$. Then, from Lemma~\ref{lem:vanderwoude}, the size of a maximum linking from $\mathcal{I}_{a}$ to $Y$ in $\mathcal{G}_{F}$ equals $r$. Suppose that vertex $i$ is not saturated by some maximum linking, say $L$, from $\mathcal{I}_{a}$ to $Y$ in $\mathcal{G}_{F}$. This implies that the  size of maximum linking, namely size of linking $L$,  from  $\mathcal{I}_{a}\setminus i$ to $Y$  equals $r$, since if it were $r-1$ then $L$ should have included vertex $i$ . Consequently, $\gnrank G^{\mathcal{I}_{a}\setminus i} =r$, and hence,  $\gnrank G^{\mathcal{I}_{a}} \neq \gnrank G^{\mathcal{I}_{a}\setminus i} +1$.~$\square$

\begin{lem}\label{lem:everymaxlinking}
Let $\mathcal{I}_{a} \subseteq \mathcal{I}$ such that $i_{1} \in \mathcal{I}_{a}$. If every maximum linking from $\mathcal{I}_{a}$ to $Y$ saturates $i_{1}$, then using $\mathcal{I}_{a}$,  for almost all  realizations $(W, \tilde{B}_{a}, C, \tilde{D}_{a}) \in (W_{\omega}, \tilde{B}_{\beta^{\prime}}, C_{\gamma}, \tilde{D}_{\sigma})$, an attacker cannot attack $i_{1}$ and remain perfectly undetetcable.~$\blacksquare$ 
\end{lem}
\textit{Proof:} Let $r \in \mathbb{Z}_{+}$ be the size of  a maximum linking from $\mathcal{I}_{a}$ to  $Y$ in the directed graph $\mathcal{G}_{F}$. Denote by $G_{L_{1}}^{\mathcal{I}_{a}}(z)$ the structured transfer function matrix.\\
Pick a maximum linking, say $L_{1}$, from $\mathcal{I}_{a}$ to  $Y$ in the directed graph $\mathcal{G}_{F}$, and do the following: for all edges of $\mathcal{G}_{F}$ that are contained in $L_{1}$, set the weights to $1$; for the rest, set the weights to $0$.  As a consequence of this choice of edge weights, and since size of  a maximum linking from $\mathcal{I}_{a}$ to  $Y$ in the directed graph $\mathcal{G}_{F}$ equals $r$, from the definition of linking it follows that for the aforementioned choice of free parameters, $\rank G_{L_{1}}^{\mathcal{I}_{a}}(z) =r$. Moreover, column $i_{1}$ is linearly independent  of the remaining columns in $G_{L_{1}}^{\mathcal{I}_{a}}(z)$. Since there exists one realization for which   $\rank G_{L_{j}}^{\mathcal{I}_{a}}(z) =r$, then for almost all choices of edge weights in $\mathcal{G}_{F}$  $\rank G^{\mathcal{I}_{a}}(z) =r$, and all such realizations would have column $i_{1}$ to be linearly independent  of the remaining columns in $G_{L_{i}}^{\mathcal{I}_{a}}(z)$. \\
\noindent Given that every maximum linking from $\mathcal{I}_{a}$ to $Y$ saturates $i_{1}$, it follows that, even though repeating the procedure above for different maximum linkings in $\mathcal{G}_{F}$ would yield different 
 $G_{L_{i}}^{\mathcal{I}_{a}}(z)$, for \emph{all} such  structured transfer function matrices, $G_{L_{i}}^{\mathcal{I}_{a}}(z)$, the following are satisfied: a) $\gnrank G_{L_{i}}^{\mathcal{I}_{a}}(z) =r$, and b) for almost all realizations, column $i_{1}$ will be linearly independent  of the remaining columns in $G_{L_{i}}^{\mathcal{I}_{a}}(z)$, $i \neq 1$.\\
Therefore, for almost all  realizations $(W, \tilde{B}_{a}, C, \tilde{D}_{a}) \in (W_{\omega}, \tilde{B}_{\beta^{\prime}}, C_{\gamma}, \tilde{D}_{\sigma})$, the following is satisfied: let $x(0) =0$, $u=0$, then $y=0$ $\implies$ $a^{(i_{1})} =0$, which, with respect to set $\mathcal{I}_{a}$, is in violation of constraint (C4). Hence, using $\mathcal{I}_{a}$,  for almost all  realizations $(W, \tilde{B}_{a}, C, \tilde{D}_{a}) \in (W_{\omega}, \tilde{B}_{\beta^{\prime}}, C_{\gamma}, \tilde{D}_{\sigma})$, an attacker cannot attack $i_{1}$ and remain perfectly undetectable.~$\square$

With Lemmas~\ref{lem:every:max:linking} and~\ref{lem:everymaxlinking} in place, the following proposition gives an upper bound for the generic actuator security index.

\begin{prop}\label{prop:suff:less:restrictive}If for every $\mathcal{I}_{a} \subseteq \mathcal{I}$ such that a) $\lvert\mathcal{I}_{a}\rvert = p$, and b) $i_{1} \in \mathcal{I}_{a}$, \[\gnrank G^{\mathcal{I}_{a}} \neq \gnrank G^{\mathcal{I}_{a}\setminus i_{1}},\] then  $\delta_{s}(i_{1}) \geq p+1$.~$\blacksquare$
\end{prop}
\textit{Proof:} Suppose that  the conditions in Prop.~\ref{prop:suff:less:restrictive} are satisfied, yet  $\delta_{s}(i_{1}) \leq p$. That is, there exists some $\mathcal{I}_{a}^{1} \subseteq \mathcal{I}$, where a) $\lvert\mathcal{I}_{a}^{1}\rvert = p$, and b) $i_{1} \in \mathcal{I}_{a}^{1}$, such that for almost all realizations $(W, \tilde{B}_{a}, C, \tilde{D}_{a}) \in \{W_{\omega}, \tilde{B}_{\beta^{\prime}}, C_{\gamma}, \tilde{D}_{\sigma}\}$, where $[\begin{smallmatrix}\tilde{B}_{\beta^{\prime}} \\ \tilde{D}_{\sigma}\end{smallmatrix}]$ is the structured submatrix of $[\begin{smallmatrix}B_{\beta^{\prime}} \\ D_{\sigma}\end{smallmatrix}]$ with column indices corresponding to elements in $\mathcal{I}_{a}^{1}$, the following is satisfied: let $x(0) =0$, $u=0$, then $y=0$ $\implies$ $a^{(i_{1})} \neq 0$.



Let $r \in \mathbb{Z}_{+}$, where $1 \leq r \leq \lvert\mathcal{I}_{a}^{1}\rvert -1$, denote the size of a maximum linking from $\mathcal{I}_{a}^{1}$ to $Y$ in $\mathcal{G}_{F}$. Then, from Lemma~\ref{lem:vanderwoude}, $\gnrank G^{{\mathcal{I}_{a}^{1}}} =r$. Hence, $\gnrank G^{\mathcal{I}_{a}^{1}\setminus i_{1}}\geq r-1$. Now there are two cases to consider here.\\
\textbf{Case a:} If $\gnrank G^{\mathcal{I}_{a}^{1}\setminus i_{1}} = r$, then, by contraposition, the proof is complete.\\
\textbf{Case b:} If $\gnrank G^{\mathcal{I}_{a}^{1}\setminus i_{1}} = r-1$, then \break $\gnrank G^{{\mathcal{I}_{a}^{1}}} = \gnrank G^{\mathcal{I}_{a}^{1}\setminus i_{1}} +1$.  Hence, from Lemma~\ref{lem:every:max:linking}, vertex $i_{1}$ is covered by every maximum linking from $\mathcal{I}_{a}^{1}$ to $Y$. This implies, from  Lemma~\ref{lem:everymaxlinking}, that for almost all realizations $(W, \tilde{B}_{a}, C, \tilde{D}_{a}) \in \{W_{\omega}, \tilde{B}_{\beta^{\prime}}, C_{\gamma}, \tilde{D}_{\sigma}\}$,  an attacker cannot attack $i_{1}$ and remain perfectly undetectable.  This contradicts our initial assumption. Thus, the proof is complete.~$\square$

 It is interesting to ask whether  the bound in Prop.~\ref{prop:suff:less:restrictive} is tight; i.e, if for some  $i \in \mathcal{I}$ $\delta_{s}(i) \geq p+1$, then is it also true that for the same $i \in \mathcal{I}$
$\delta_{s}(i) \leq p+1$?
The following proposition answers this question.
\begin{prop} \label{prop:nece}
Let  $\mathcal{I}_{a} \subseteq \mathcal{I}$, and $i_{1} \in \mathcal{I}_{a}$. If  \[ \gnrank G^{\mathcal{I}_{a}} = \gnrank G^{\mathcal{I}_{a}\setminus i_{1}},\] then $\delta_{s}(i_{1}) \leq \lvert\mathcal{I}_{a}\rvert$.~$\blacksquare$
\end{prop}
\textit{Proof:}  Let us denote by $[\begin{smallmatrix} \hat{B}_{\beta^{\prime}} \\ \hat{D}_{\sigma}\end{smallmatrix}]$ the columns of $[\begin{smallmatrix} B_{\beta^{\prime}} \\ D_{\sigma}\end{smallmatrix}]$ that correspond to elements in $\mathcal{I}_{a}$. By assumption, all such structured submatrices $[\begin{smallmatrix} \hat{B}_{\beta^{\prime}} \\ \hat{D}_{\sigma}\end{smallmatrix}]$ contain column $i_{1}$. Let us denote by $(W_{\omega}, \hat{B}_{\beta^{\prime}}, C_{\gamma}, \hat{D}_{\sigma})$ the corresponding structured subsystem of $(W_{\omega}, B_{\beta^{\prime}}, C_{\gamma}, D_{\sigma})$. Let $G^{\mathcal{I}_{a}}$ denote the corresponding structured transfer function matrix, and let $\gnrank G^{\mathcal{I}_{a}} =r$, where $1\leq r \leq  \lvert\mathcal{I}_{a}\rvert$. \\ 
Suppose that $\delta_{s}(i_{1}) > \lvert\mathcal{I}_{a}\rvert$. Then, together with the   definition of actuator security indices, it follows that for every set of size $\lvert\mathcal{I}_{a}\rvert$ containing $i_{1}$, for almost all realizations $(W, \hat{B}_{a}, C, \hat{D}_{a}) \in (W_{\omega}, \hat{B}_{\beta^{\prime}}, C_{\gamma}, \hat{D}_{\sigma})$, the following is satisfied: let $x(0) =0$, $u=0$, then $y=0$ $\implies$ $a^{(i_{1})} =0$. This, since $G^{\mathcal{I}_{a} }= \begin{bmatrix}G^{\mathcal{I}_{a}\setminus {\{i_{1}\}}} & G^{\{i_{1}\}} \end{bmatrix}$, implies that  $\gnrank G^{\{i_{1}\}}=1$, and moreover,  column $i_{1}$ is linearly independent of the columns of $G^{\mathcal{I}_{a}\setminus {\{i_{1}\}}}$. Therefore, $\gnrank G^{\mathcal{I}_{a}\setminus {\{i_{1}\}}} = r-1$, which further implies that $\gnrank G^{\mathcal{I}_{a} } \neq \gnrank G^{\mathcal{I}_{a}\setminus {\{i_{1}\}}}$. This completes the proof.~$\square$


\noindent Combining the results in Prop.~\ref{prop:suff:less:restrictive} and~\ref{prop:nece}, yields the following.
\begin{prop} \label{key:result:algebraic} Let $p \in \mathbb{Z}_{+}$ such that $1 \leq p \leq \lvert\mathcal{I}\rvert$.
\begin{enumerate}[label=(\roman*)]
\item \label{ref:lowerbound}   If for every $\mathcal{I}_{a} \subseteq \mathcal{I}$ such that a) $\lvert\mathcal{I}_{a}\rvert = p$, and b) $i_{1} \in \mathcal{I}_{a}$, $\gnrank G^{\mathcal{I}_{a}} \neq \gnrank G^{\mathcal{I}_{a}\setminus i_{1}}$, then  $\delta_{s}(i_{1}) \geq p+1$.
\item \label{ref:upperbound}  If $\gnrank G^{\mathcal{I}_{a}} = \gnrank G^{\mathcal{I}_{a}\setminus i_{1}}$, then  $\delta_{s}(i_{1}) \leq p$.~$\blacksquare$
\end{enumerate}
\end{prop}



 Note that the conditions in Prop.~\ref{key:result:algebraic} give lower bound and upper bound on the number of elements needed to conduct an attack using a particular actuator (or sensor) for almost all choices of edge weights in $\mathcal{G}_{F}$. It turns out that the \emph{exact} number of elements needed  is generic, i.e., in order to conduct an attack using a particular component for almost all choices of edge weights in $\mathcal{G}_{F}$, the same number of elements are needed. The following theorem addresses this.
\begin{thm} \label{ref:main-result1}
For an actuator $i \in \mathcal{I}$, the  security index, $\delta(i)$, is generic.~$\blacksquare$
\end{thm}
\textit{Proof:} First note that the conditions in Prop.~\ref{key:result:algebraic} are mutually exclusive. For $p =1$, thanks to Assumption~\ref{assm:dummy},  the condition in item~\ref{ref:lowerbound}  is trivially satisfied for every $i \in \mathcal{I}$. Therefore, one needs to check for $p \neq 1$, starting from $p=2$. This is done as follows: generate all sets $\mathcal{I}_{a}$ such that $i \in \mathcal{I}_{a}$ and $\lvert\mathcal{I}_{a}\rvert = p$. If the condition in item~\ref{ref:lowerbound} is satisfied for every such set $\mathcal{I}_{a}$, then at least $p+1$ elements are needed. Let $\mathcal{I}_{a}$ be the first set for which the condition in item~\ref{ref:upperbound} (resp. item~\ref{ref:lowerbound}) is satisfied (resp. violated). 
Then a perfectly undetectable attack using actuator $i$ can be conducted with at most $p$ elements for almost all choices of edge weights in $\mathcal{G}_{F}$. However, since the condition was satisfied for $p=p-1$ ,  it follows that in order to conduct an attack using actuator $i$ \emph{exactly} $p$ elements are needed for almost all choices of edge weights in $\mathcal{G}_{F}$. This implies that  $\delta_{s}(i) = p$.~$\square$


\noindent \seb{Theorem~\ref{ref:main-result1} answers question~\ref{q1} in Sect.~\ref{pb:stmnt}, and should be interpreted in the following sense: For almost all choices of free parameters in $\mathbb{R}^{\alpha}$, the security index of actuator $i$, $\delta(i)$, remains the same (i.e., $p$). For some choices of free parameters in $\mathbb{R}^{\alpha}$, $\delta(i)$ could be different from $p$. However, all such choices of free parameters would lie on a set of Lebesgue measure zero in the space of free parameters. }

 In the context of network systems, one is often interested in graph-theoretic conditions as opposed to algebraic ones. This motivates us to seek graph-theoretic alternative  for the conditions given in Prop.~\ref{key:result:algebraic}. The following lemma eases this transition.
\begin{lem} \label{key:graphlemma} Let $\mathcal{I}_{a} \subseteq \mathcal{I}$ such that  $i_{1} \in \mathcal{I}_{a}$, and let $\mathcal{G}_{F}$ be the associated directed (sub)graph. Vertex $i_{1}$ is saturated by every maximum linking from $\mathcal{I}_{a}$ to $Y$ in $\mathcal{G}_{F}$ if and only if $\gnrank G^{\mathcal{I}_{a}} = \gnrank  G^{\mathcal{I}_{a}\setminus i_{1}} +1$.~$\blacksquare$
\end{lem}
\textit{Proof:} Let $r$, where $1 \leq r \leq \lvert\mathcal{I}_{a}\rvert$ be the size of a maximum linking from $\mathcal{I}_{a}$ to $Y$ in $\mathcal{G}_{F}$. Then, from Lemma~\ref{lem:vanderwoude},
$\gnrank G^{\mathcal{I}_{a}} =r$. Suppose $i_{1}$ is saturated by every maximum linking from $\mathcal{I}_{a}$ to $Y$ in $\mathcal{G}_{F}$, then  removing vertex $i_{1}$ from $\mathcal{I}_{a}$ should reduce the size of the maximum linking from  $\mathcal{I}_{a}\setminus i_{1}$ to $Y$ in $\mathcal{G}_{F}$ by one, because otherwise there would be a maximum linking  from $\mathcal{I}_{a}$ to $Y$ in $\mathcal{G}_{F}$ that does not saturate $i_{1}$. Hence, the maximum size of a  linking from  $\mathcal{I}_{a}\setminus i_{1}$ to $Y$ equals $r-1$. This further implies that  $\gnrank  G^{\mathcal{I}_{a}\setminus i_{1}} = r-1$.\\
The other direction is handled in Lemma~\ref{lem:every:max:linking}.~$\square$

 From Prop.~\ref{key:result:algebraic} and Lemma~\ref{key:graphlemma}, the following is readily obtained.
\begin{thm}[Graph-theoretic conditions] \label{ref:main-result2} Let $p \in \mathbb{Z}_{+}$ such that $1 \leq p \leq \lvert\mathcal{I} \rvert$. Then
\begin{enumerate}[label=(\roman*)]
\item \label{graph:cond1} If for every $\mathcal{I}_{a} \subseteq \mathcal{I}$ such that a) $i_{1} \in \mathcal{I}_{a}$, and b)$\lvert\mathcal{I}_{a} \rvert =p$, $i_{1}$ is saturated by every maximum linking from $\mathcal{I}_{a}$ to $Y$ in $\mathcal{G}_{F}$, then  $\delta_{s}(i_{1}) \geq p+1$;
\item \label{graph:cond2} If for some $\mathcal{I}_{a} \subseteq \mathcal{I}$ such that a) $i_{1} \in \mathcal{I}_{a}$, and b)$\lvert\mathcal{I}_{a} \rvert =p$, there exists a maximum linking from $\mathcal{I}_{a}$ to $Y$ in $\mathcal{G}_{F}$ that does not saturate $i_{1}$, then $\delta_{s}(i_{1}) \leq p$.~$\blacksquare$
\end{enumerate}
\end{thm}
Theorem~\ref{ref:main-result2} answers question~\ref{q2} in Sect.~\ref{pb:stmnt}.\\
Some special cases are of interest; these are addressed by the following remarks. 
\begin{rem}\label{rem:p=max}
If condition in item~\ref{graph:cond1}  of Theorem~\ref{ref:main-result2} is satisfied for the particular case of $p = \lvert \mathcal{I}\rvert$, then for almost all choices of edge weights in $\mathcal{G}_{F}$ the attacker cannot attack $i$ while staying perfectly undetectable. Hence,  $\delta_{s}(i) = +\infty$.~$\blacksquare$
\end{rem}

In the context of perfectly undetectable attacks, a closely-related  notion is that of left-invertibility. A system is left-invertible if given a sequence of outputs and initial state, the unknown inputs, in our case attacks, can be uniquely recovered up to some delay. The next two remarks show the interplay between generic left-invertibility and the conditions in Theorem~\ref{ref:main-result2}.


\begin{rem}\label{rem:counterexample}
The condition in Remark~\ref{rem:p=max} being satisfied does not necessarily imply that the system $(W_{\omega}, B_{\beta^{\prime}}, C_{\gamma}, D_{\sigma})$ is generically left-invertible. To see this, consider the  example in Figure~\ref{fig:1}.
\begin{figure}[h!]
\centering
\includegraphics[height=0.25\columnwidth]{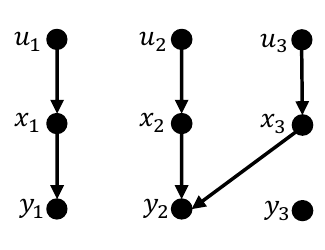}
\caption{ $\mathcal{G}_{2}$}
\label{fig:1}
\end{figure}
Let $u_{1}$ be the vertex of interest, $p=3$, and, therefore, $\mathcal{I}_{a} = \{u_{1}, u_{2}, u_{3}\}$. The list of all  maximum linkings are as follows: $\{(u_{1}\rightarrow x_{1} \rightarrow y_{1}), (u_{2}\rightarrow x_{2} \rightarrow y_{2})\}$; and $\{(u_{1}\rightarrow x_{1} \rightarrow y_{1}), (u_{3}\rightarrow x_{2} \rightarrow y_{2})\}$. Note that $u_{1}$ is saturated by both the maximum linkings. Hence,  $\delta_{s}(u_{1}) = + \infty$. However, since maximum size of a linking from $\mathcal{I}_{a}$ to $Y$ is $2$, from Theorem~C.5 in \cite{sundaram2012fault}, the structured system represented by graph $\mathcal{G}_{2}$ is generically not left-invertible.~$\blacksquare$
\end{rem}

\begin{rem}
Suppose that the condition in  item~\ref{graph:cond1}  of Theorem~\ref{ref:main-result2} is satisfied for the particular case of $p = \lvert \mathcal{I}\rvert$ for \emph{every} $i \in \mathcal{I}$, then, from Remark~\ref{rem:p=max}, $\delta_{s}(i) = + \infty$ for every $i \in \mathcal{I}$. This also implies that the maximum size of a linking from $\mathcal{I}$ to $Y$ in $\tilde{\mathcal{G}}$ is $\lvert \mathcal{I}\rvert$. Hence, the system $(W, B_{a}, C, D_{a})$ is generically left-invertible.~$\blacksquare$
\end{rem}
\noindent Note that the condition~\ref{graph:cond1} in Theorem~\ref{ref:main-result2} asks that for every $\mathcal{I}_{a}$ that satisfies the given criteria, the actuator $i \in V_{ess}(\mathcal{I}_{a}, \mathcal{Y})$. For a given $\mathcal{I}_{a}$, whether  $i \in V_{ess}(\mathcal{I}_{a}, \mathcal{Y})$   can be checked efficiently using depth search algorithms \cite{boukhobza2007state}. With this understanding in place, we can now present an approach for computing  $\delta_{s}(i)$.
\subsection*{Calculating structural security index} \label{sect:algo}

In this subsection, we introduce a brute force search method for calculating a structural security index $\delta_s(i)$ of an actuator $i$. The idea is to iterate through all subsets $\mathcal{I}_{a}$ that contain $i$ and have cardinality $p$, $1 \leq p \leq |\mathcal{I}|$, starting from $p=1$. If we find a subset for which condition~\ref{graph:cond2} in Theorem~\ref{ref:main-result2} is satisfied, then $\delta_{s}(i) = p$, and we stop the search. Otherwise, we increment $p$,  and repeat the process. 

\begin{algorithm} [h] \label{algorithm1}
\caption{Calculating the structural security index $\delta_s(i)$ of an actuator $i$. }
\begin{algorithmic}
\State \textbf{Input:} $W_{\omega}$,$B_{\beta^{\prime}}$,$C_{\gamma}$,$D_{\sigma}$,   $i \in \mathcal{I}$ 
\State \textbf{Output:} $\delta_s(i)$
\State Step~0: Set $p=1$ ; 
\State Step~1: Generate all subsets $\mathcal{I}_a$ of $\mathcal{I}$ such that: (i)~$|\mathcal{I}_a|=p$; (ii)~$i \in \mathcal{I}_a$; and calculate $q$, the number of such subsets
\State Step~2: for j=1:q
\State Step~3: compute $V_{ess}(\mathcal{I}_{a}^{j}, \mathcal{Y})$ in $\tilde{\mathcal{G}}$\\
- if $i \in V_{ess}(\mathcal{I}_{a}^{j}, \mathcal{Y}) $, then $j=j+1$, and go to step 3
-  else $\delta_{s}(i) =p$, and terminate the algorithm.

\State Step~4: If $p \leq \lvert\mathcal{I}\rvert$, then $p=p+1$, and return to step 1\\
-else return $\delta_s(i)= + \infty$, and terminate the algorithm. 
\end{algorithmic}
\label{algorithm:impact-calculation}
\end{algorithm}


 With respect to the example in Figure~\ref{fig:running:example}, $\mathcal{I}=\{u_{1}, u_{2}, a_{y_{1}}\}$, 
 For vertex, $u_{1}$, consider the vertex set $\mathcal{I}_{a} = \{u_{1}, a_{y_{1}}\}$. The maximum size of a linking from $\mathcal{I}_{a}$ to $Y$ is $1$; there are two such linkings, namely: i) $ u_{1} \rightarrow x_{1} \rightarrow y_{1}$ and ii) $a_{y_{1}} \rightarrow y_{1}$. Note that $V_{ess}(\mathcal{I}_{a}, Y) =\{y_{1}\}$, and, therefore, $u_{1} \notin V_{ess}(\mathcal{I}_{a}, Y)$. Hence, $\delta_{s}(u_{1}) = 2$. It can also be seen that $\delta_{s}(a_{y_{1}}) = 2$.\\
To compute $\delta_{s}(u_{2})$, consider the vertex set $\mathcal{I}=\{u_{1}, u_{2}, a_{y_{1}}\}$. The maximum linkings from $\mathcal{I}$ to $Y$ are the following:
\begin{itemize}

\item $(u_{1} \rightarrow x_{1} \rightarrow y_{1}, u_{2} \rightarrow x_{2} \rightarrow y_{2})$; 

\item $(u_{1} \rightarrow x_{1} \rightarrow y_{1},  u_{2} \rightarrow x_{2} \rightarrow x_{3} \rightarrow y_{3})$;

\item  $(u_{1} \rightarrow x_{1} \rightarrow y_{1},  u_{2} \rightarrow x_{2} \rightarrow x_{3} \rightarrow x_{4} \rightarrow y_{3})$; \item $(a_{y_{1}} \rightarrow y_{1},  u_{2} \rightarrow x_{2} \rightarrow y_{2})$;

\item $(a_{y_{1}} \rightarrow y_{1}, u_{2} \rightarrow x_{2} \rightarrow x_{3} \rightarrow y_{3})$; and

\item $(a_{y_{1}} \rightarrow y_{1}, u_{2} \rightarrow x_{2} \rightarrow x_{3} \rightarrow x_{4} \rightarrow y_{3})$
\end{itemize}
It can be immediately seen that $u_{2} \in V_{ess}(\mathcal{I}, Y)$, and, hence from Remark~\ref{rem:p=max} $\delta_{s}(u_{2}) = + \infty$.

\section{Conclusion} \label{sect:conclusion}
This paper tackled the problem of computing actuator security indices for discrete-time LTI network systems.  First, we showed that, under a structured systems framework, the actuator security index is generic. Subsequently, we provided graph-theoretic conditions for computing the structural actuator security index. Based on those conditions, we presented an algorithm for the same.\\
  
In the future work, we will focus on computing the index $\delta_s$ efficiently and developing defense strategies for improving this index. 


\bibliography{ReferencesKTH}
\end{document}